\documentclass{desyproc}

\newcommand{\JD}{\mathrm{JD}}

\newcommand{\lp}{\left(}
\newcommand{\rp}{\right)}

\usepackage{cite}

\begin{document}
\title{Progress in jet reconstruction and heavy ion collisions}

\author{{\slshape Juan Rojo}\\[1ex]
INFN, Sezione di Milano, Via Celoria 16, I - 20133, Milano (Italy)}

\contribID{rojo\_juan}

\desyproc{DESY-PROC-2009-xx}
\acronym{EDS'09} 
\doi  

\maketitle

\begin{abstract}
We review recent developements related to jet clustering
algorithms and jet reconstruction, with particular
emphasis on their implications in heavy ion collisions. 
These developements include fast implementations
of sequential recombination algorithms,
 new IRC safe algorithms, quantitative
determination of jet areas and
quality measures for jet finding, among many others.
We also show how jet reconstruction provides a useful tool to
probe the characteristics of the hot and dense medium created
in heavy ion collisions, which allows one to distinguish between different
models of parton-medium interaction.
\end{abstract}

\paragraph{Recent developements in jet algorithms}
With the upcoming start-up of the proton-proton and
heavy ion programs at the LHC,  
jet reconstruction techniques have been the subject
of intense research in the recent 
years~\cite{Ellis:2007ib,Buttar:2008jx,Salam:2009jx,Cacciari:2009fi}.
In this contribution we briefly review this progress with
special emphasis on their implications in heavy ion collisions.

An important developement  has been  the
fast implementation of the $k_T$~\cite{Catani:1993hr}
and Cambridge/Aachen~\cite{Catani:1991hj,Wobisch:1998wt} 
jet algorithms.
Prior to 2005, existing
implementations scaled as $N^3$, with $N$ the number
of particles to be clustered, thus making it unpractical
for high multiplicity proton-proton collisions, 
and even more in heavy ion collisions (HIC).
Thanks to computational geometry methods, the performance
of these algorithms was made to scale as $N\ln N$~\cite{Cacciari:2005hq}. 
These fast implementations are available through 
the {\tt FastJet} package~\cite{fastjet}, together
with area-based subtraction methods and plugins to external
jet finders.

Another important achievement has been
the formulation of a practical (scaling as $N^2 \ln N$)
 infrared and collinear (IRC) safe
 cone algorithm, SISCone~\cite{Salam:2007xv}. 
Unlike all other commonly used
cone algorithms, SISCone is IRC safe
to all orders in perturbation theory by construction. 
This property allows one to compare any perturbative computation
with experimental data, which for IRC unsafe algorithms
is impossible beyond some fixed 
order~\cite{Salam:2007xv}. The phenomenological
implications of SISCone when compared with the (IRC unsafe)
 commonly used MidPoint cone algorithm range from  few percent
differences in the inclusive jet spectrum, somewhat larger
in the presence of realistic Underlying Event (UE), up to 50\% differences
for more exclusive observables, like jet-mass spectra
in multi-jet events.

Another recently developed IRC safe
jet algorithm is the anti-$k_t$ 
algorithm~\cite{Cacciari:2008gp}. This algorithm
is related to  $k_T$ and Cam/Aa  by its
distance measure,  $  d_{ij} \equiv { \min(k_{ti}^{ 2p 
},k_{tj}^{2p })} 
{ \Delta R_{ij}^2/R^2}$,
with $p=-1$ ($p=1$ corresponds to $k_T$ and $p=0$ to Cam/Aa).
The anti-$k_T$ algorithm  has the property of
being soft-resilient, that is,
due to its distance soft particles are always clustered
with hard particles first.
This property leads to rather regular jet areas,
which become perfectly circular in the limit in which all
hard particles are separated in the $(y,\phi)$ plane
by at least a distance $2R$~\cite{Cacciari:2008gp}.
Another important advantage of the anti-$k_t$  algorithm is
that it has a very small back-reaction,
that is, the presence of a soft background has reduced effects
on which hard particles are clustered into a given jet. This is
a particularly relevant advantage of the anti-$k_t$ algorithm
for jet reconstruction in very dense environments like
heavy ion collisions.

There has been historically some confusion  about the
concept of the {\it area} of a jet, specially since the
naive expectation $A_{\rm jet}=\pi R^2$
only holds at leading order. The situation  was recently clarified by the
introduction of quantitative definitions of jet areas
based on the {\it catchment} properties of hard jets with respect 
to very soft
particles, called {\it ghosts} in~\cite{Cacciari:2008gn}. 
On top of their
theoretical interest, jet areas have important
applications related to the subtraction of
soft backgrounds coming from the UE or from Pile-Up (PU),
 both in proton-proton and in
heavy-ion collisions, as discussed in~\cite{Cacciari:2007fd}.

\paragraph{Performance of jet algorithms}
A recurring question in jet studies is ``what is the best jet
definition for a given specific analysis under certain 
experimental conditions''? 
Most existing techniques either 
use as a reference  unphysical Monte Carlo partons (an ambiguous
concept beyond LO)
and/or assume some shape for the measured kinematical distributions, 
typically a gaussian.
To overcome these disadvantages, 
a new strategy to quantify the performance of jet definitions in
kinematic reconstruction tasks has been recently
introduced~\cite{jet-performance},  which was designed to make use exclusively
of physical observables. Related studies which
address the same question were discussed in 
Ref.~\cite{Dasgupta:2007wa,Krohn:2009zg}.

In Ref.~\cite{jet-performance} two quality measures respecting
the above requirements are proposed, and applied to the kinematic
reconstruction of invariant mass distributions in dijet events
from hadronically decaying heavy resonances
in simulated LHC proton-proton collisions
for a wide range of energies.
These quality measures can in turn be mapped into an effective luminosity
ratio, defined as
\begin{equation}
  \label{eq:rhol_basic_def}
  \rho_{\cal L}(\JD_2 / \JD_1) \equiv 
  \frac{{\cal L}(\text{needed with }\JD_2)}
       {{\cal L}(\text{needed with }\JD_1)} 
  = \left[
    \frac{\Sigma\lp \JD_1 \rp} {\Sigma\lp \JD_2 \rp} \right]^2 \, .
\end{equation}
Given a certain signal significance $\Sigma$ with 
jet definition $\JD_2$, $\rho_{\cal
  L}(\JD_2/\JD_1)$ indicates the factor more luminosity needed to
obtain the same significance as with jet definition $\JD_1$.

\begin{figure}
\begin{center}
\includegraphics[width=0.85\textwidth]{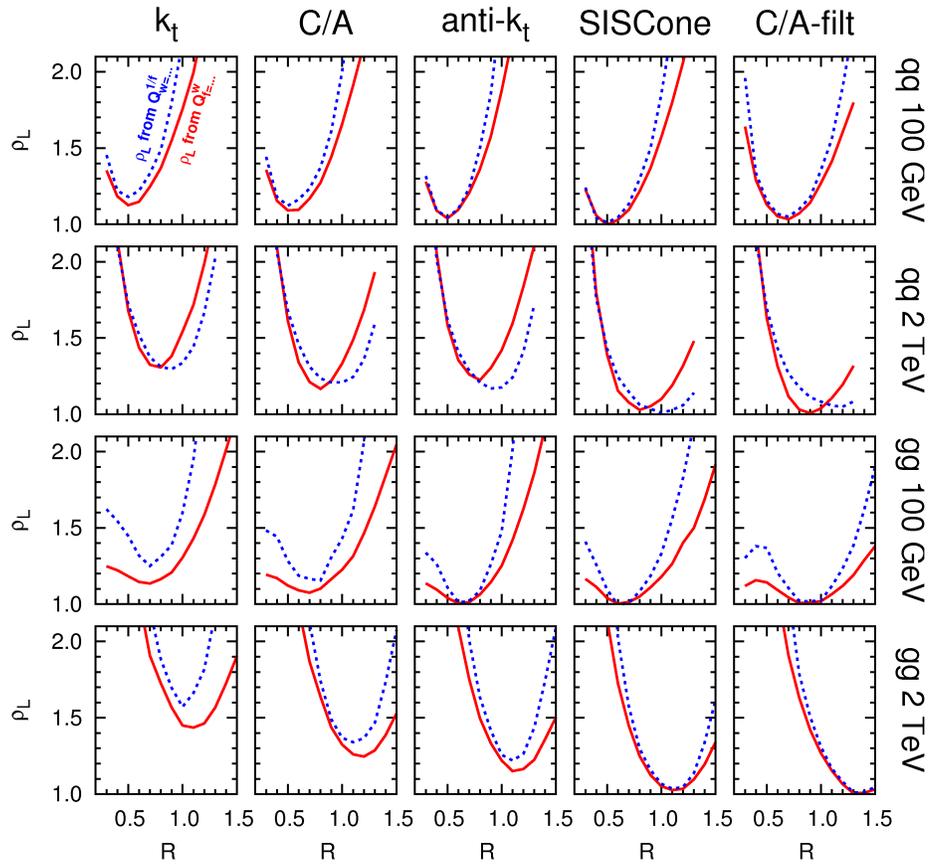}
\caption{\small 
The effective luminosity ratio, 
Eq.~\ref{eq:rhol_basic_def}, for quark and gluon jets
at 100 GeV and 2 TeV, from Ref.~\cite{jet-performance}.
\label{fig:rhoL-summary} }
\end{center}
\end{figure}

The results of~\cite{jet-performance}
 over a large range
of jet definitions,\footnote{There results can also be accessed through an
interactive web tool~\cite{jet-performance-web} which allows
the user to compare the jet finding quality for a wide range of 
parameters (jet algorithm, $R$, value of PU, ...).} 
summarized in Fig..~\ref{fig:rhoL-summary},
indicate that for gluon jets, and in general for TeV scales, there
are significant benefits to be had from using larger radii that
those commonly used, up to
$R\gtrsim 1$, while smaller radii are favored for smaller
values of the jet $p_T$. 
In general, SISCone and C/A-filt (Cam/Aa supplemented
with a filtering procedure~\cite{Butterworth:2008iy}) show the
best performance. These conclusions are robust in the presence
of high-luminosity PU, when subtracted with the
jet area technique~\cite{Cacciari:2007fd}. 

The same philosophy 
could be applied to heavy ion collisions to determine how 
in such case
the optimal jet definition is affected by the overwhelming
underlying event present. It is clear however that
 in this case the quality
measures are different from the proton-proton case: for example,
for HIC resonance reconstruction is not a relevant quality measure.

\paragraph{Jet reconstruction in heavy ion collisions}
While QCD jets are ubiquitous in pp collisions,
until recently \cite{Lai:2009zq,Salur:2009vz} 
no real jet reconstruction had been obtained
 in the much more
challenging environment of HIC. Indeed, usually in HIC one refers to
a single hard particle in the event as a {\it jet}. However, 
reconstructing full QCD
jets provides a much more precise window to the
properties of the hot and dense
medium created in the collision than just leading
particles.

\begin{figure}
\begin{center}
\includegraphics[width=0.55\textwidth]{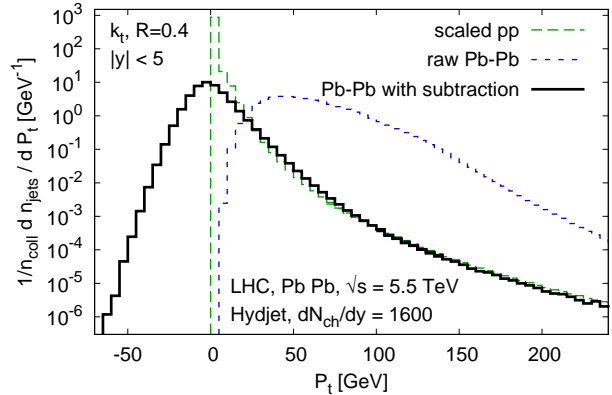}
\caption{\small The simulated inclusive jet spectrum at the
LHC with the $k_T$ algorithm, including subtraction,
from Ref.~\cite{Cacciari:2007fd}.}
\label{fig:plot-hic}
\end{center}
\end{figure}

The difficulty in
reconstructing jets in HIC stems from the huge backgrounds,
which need to be subtracted in order to compare with baseline results, like
proton-proton or proton-ion collisions at the same energy.
There are various techniques to subtract such large backgrounds. 
In Ref.~\cite{Cacciari:2007fd} it was shown 
how the area method introduced above could efficiently subtract 
large UE backgrounds in HIC for LHC conditions with good accuracy, 
see Fig.~\ref{fig:plot-hic}.

It is therefore important for precision measurements 
to  control the accuracy of the
subtraction procedure intrinsic to jet reconstruction
in HIC, as well
as to understand the differences between the performances
of different jet
algorithms. Ongoing studies~\cite{hic}
suggest that one of the dominant
sources of systematic error in HIC jet reconstruction
is back-reaction~\cite{Cacciari:2008gn}, therefore anti-$k_t$ is potentially
interesting in this situation due to its small 
back-reaction~\cite{Cacciari:2008gp}. 
Ref.~\cite{hic} also investigates
the use of local ranges for the
determination of the background level $\rho$ as a technique
to reduce the 
effects of point-to-point background fluctuations.

Related to progress in jet reconstruction algorithms, an important
developement in recent years has been the developement
of several exclusive Monte Carlo event generators for
heavy ion collisions which account for
the interaction of partons propagating within the hot and dense
medium created in the
 collision~\cite{Armesto:2009fj,Zapp:2008gi,Lokhtin:2008xi,Armesto:2009ab}. 
These event generators, which assume different models for the
parton-medium interaction~\cite{Salgado:2009jp}, can be used together with 
modern jet reconstruction techniques in order
to determine, under realistic experimental conditions, which 
jet-related observables are at the same time more robust and
more sensitive to the different scenarios for the hot and dense medium
dynamics.

\begin{figure}
\begin{center}
\includegraphics[width=0.80\textwidth]{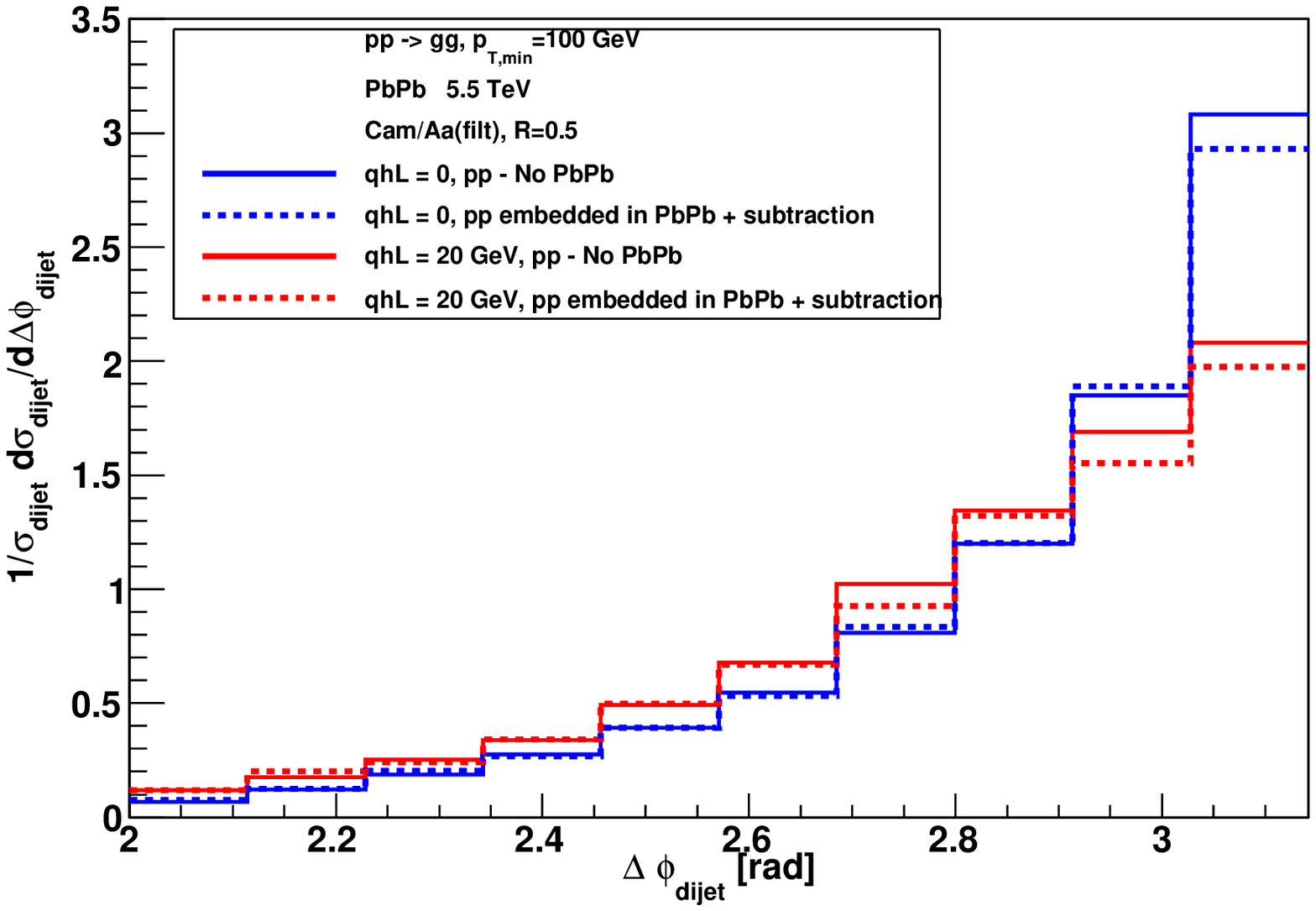}
\includegraphics[width=0.80\textwidth]{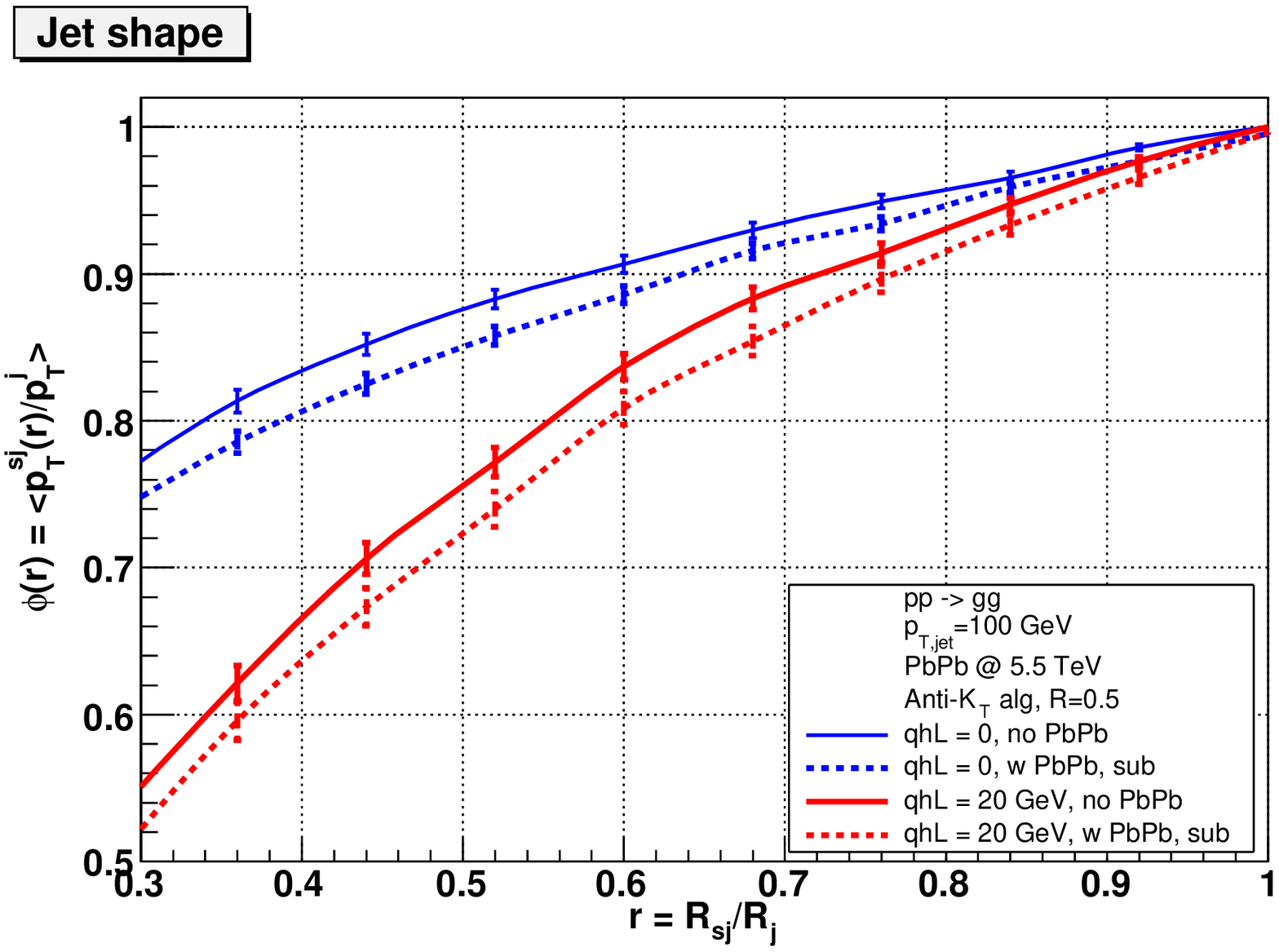}
\caption{\small Preliminary results for full jet
reconstruction, including background subtraction,
of jet-related observables
 under realistic experimental conditions at the
LHC: dijet azimuthal correlations (upper plot) and
jet shapes (lower plot). See text for details.}
\label{fig:plot-hi-obs}
\end{center}
\end{figure}

As an illustration of how these Monte Carlo programs with
medium effects can be coupled to jet reconstruction
techniques and be used to determine medium properties
in HIC, 
in Fig.~\ref{fig:plot-hi-obs} we show preliminary results for
two observables which are sensitive to medium effects:
the dijet azimuthal correlations and the jet shape.
Jet shapes are defined analogously to~\cite{Armesto:2009fj}.
To obtain these results,
hard events with $p_T^{\rm min} = 100$ GeV 
are generated and then propagated through 
a model of the
hot medium by means of the {\tt Q-PYTHIA} 
Monte Carlo~\cite{Armesto:2009fj}, for
different values of the medium parameters. 
The resulting hadronic event is embedded into a minimum bias
PbPb event generated with the PSM Monte Carlo~\cite{Amelin:2001sk}
for different scenarios of central multiplicity at the LHC.
Jet reconstruction is performed with different algorithms
of the {\tt FastJet} package, and UE subtraction is performed
with the jet area method. In the particular examples of
Fig.~\ref{fig:plot-hi-obs} the C/A(filt) and
anti-$k_T$ algorithms are used with the jet radius chosen
to be $R=0.5$.

For the two examples of Fig.~\ref{fig:plot-hi-obs}, we show
the proton-proton baseline results with and without
parton-medium interactions, whose strength is characterized
by the parameter $\hat{q}L$~\cite{Armesto:2009fj}, with
$L$ the medium length. The medium effects are clearly visible
for the two observables, inducing a broadening of the 
jet shape~\cite{Vitev:2008rz}, 
and a decorrelation of the dijet azimuthal spectrum.
These
curves, with no HIC UE, are labelled as {\it 'No PbPb'}. 
Note in the case of dijet correlations that dijet angles are
typically better measured than $p_T$ spectra, so
this observable is a promising candidate for an early measurement
at the LHC to characterize the hot medium.

Then we also show the
corresponding results when the pp event is embedded into
the PbPb event, in which case the UE event has been subtracted
with the area method.  These
curves are labelled as {\it 'PbPb, subtraction'}. We observe how
after the background subtraction the baseline proton-proton
results are reasonably recovered, in both cases with 
($\hat{q}L\ne 0$) and without ($\hat{q}L=0$) medium effects.

 These preliminary results 
indicate that medium-sensitive jet related observables can
be accurately reconstructed even in the presence of large
backgrounds, and are thus useful probes of the details
of the parton-medium interactions. More work however is
required to quantify the accuracy with which the hot and dense
medium created in HIC, and the values of the parameters which
characterize it, can be studied by reconstructed jets and
related observables.

\paragraph{Acknowledgments}
The author wants to acknowledge M. Cacciari, G. Salam and
G. Soyez for collaboration in this research and for
suggestions on this write-up, as well
as N. Armesto and C. Salgado for assistance with {\tt Q-PYTHIA}.


\begin{footnotesize}


\providecommand{\href}[2]{#2}\begingroup\raggedright\endgroup

\end{footnotesize}


\end{document}